\documentclass[11pt]{article}
\usepackage{amsmath,amssymb,color,graphics,epsfig,cite}

\usepackage{amsfonts}

\newcommand{\hoch}[1]{$\, ^{#1}$}

\makeatletter
\@addtoreset{equation}{section}
\makeatother

\newcommand{\be}{\begin{equation}}
\newcommand{\ee}{\end{equation}}
\newcommand{\bea}{\setlength\arraycolsep{2pt} \begin{eqnarray}}
\newcommand{\eea}{\end{eqnarray}}
\newcommand{\nn}{\nonumber}

\def\ft#1#2{{\textstyle{\frac{\scriptstyle #1}{\scriptstyle #2} } }}
\def\fft#1#2{{\frac{#1}{#2}}}

\def\0{{\sst{(0)}}}
\def\1{{\sst{(1)}}}
\def\2{{\sst{(2)}}}
\def\3{{\sst{(3)}}}
\def\4{{\sst{(4)}}}
\def\5{{\sst{(5)}}}
\def\6{{\sst{(6)}}}
\def\7{{\sst{(7)}}}
\def\8{{\sst{(8)}}}
\def\sst#1{{\scriptscriptstyle #1}}

\def\R{{\mathbb R}}

\begin{document}


\begin{center}

{\Large {\bf Butterfly Velocity Bound and Reverse Isoperimetric Inequality}}

\vspace{40pt}
{\bf Xing-Hui Feng and  H. L\"u\hoch{*}}

\vspace{10pt}

{\it Center for Advanced Quantum Studies, Department of Physics, \\
Beijing Normal University, Beijing 100875, China}

\vspace{40pt}

\underline{ABSTRACT}
\end{center}

We study the butterfly effect of the  AdS planar black holes in the framework of Einstein's general relativity. We find that the butterfly velocities can be expressed by a universal formula $v_{\rm B}^2 = TS/(2V_{\rm th} P)$. In doing so, we come upon a near-horizon geometrical formula for the thermodynamical volume $V_{\rm th}$. We verify the volume formula by examining a variety of AdS black holes.  We also show that the volume formula implies that the conjectured reverse isoperimetric inequality follows straightforwardly from the null-energy condition, for static AdS black holes.  The inequality is thus related to an upper bound of the butterfly velocities.

\vfill {\footnotesize \hoch{*}mrhonglu@gmail.com}

\thispagestyle{empty}

\pagebreak

\tableofcontents
\addtocontents{toc}{\protect\setcounter{tocdepth}{2}}



\section{Introduction}

The butterfly effect is associated with the exponential growth of a small perturbation to a quantum system. In the context of holography, this effect has a beautiful realization \cite{Shenker:2013pqa,Shenker:2013yza,Roberts:2014isa,Shenker:2014cwa,Maldacena:2015waa}
in terms of a gravitational shock wave near the horizon of an AdS black hole \cite{Sfetsos:1994xa}.  The butterfly velocity, i.e.~the velocity for the shockwave, for the Schwarzschild-AdS planar black hole in general $D$ dimensions, turns out to be constant, given by \cite{Shenker:2013pqa}
\be
v_{\rm B}^2 = \fft{D-1}{2(D-2)}\,.
\ee
The butterfly velocities for a variety of AdS planar black holes of various matter energy-momentum tensor were obtained \cite{Blake:2016wvh,Roberts:2016wdl,Blake:2016sud,Reynolds:2016pmi,
Lucas:2016yfl,Huang:2016izp,Ling:2016ibq,Ling:2016wuy}.  The study has been further generalized to include higher-order gravities \cite{Roberts:2014isa,Alishahiha:2016cjk}.  The expression of $v^{}_{\rm B}$ can be simple or complicated depending on the detail structures of the black holes.

On the other hand, based on the no-hair theorem, black holes are supposed to be the purest and hence simplest thermodynamical systems.  The physics are specified only by their thermodynamical quantities including mass and charges.  It is thus not unreasonable to expect that the butterfly velocity be a dimensionless ratio of some (dimensionful) thermodynamical quantities of the black holes.  In fact, based on this principle, it was shown that the holographic sheer viscosity-entropy ratio \cite{Policastro:2001yc,Kovtun:2003wp} is a simple consequence of the generalized Smarr relation of the thermodynamical variables \cite{Liu:2015tqa}.

In this paper, we study the butterfly effect of planar black holes in the framework of Einstein's general relativity, with only minimally coupled matter.  We limit the discussion to isotropic solutions where the toroidal (or Euclidean) plane is uniform. We find a universal formula for the butterfly velocities
\be
v^2_{\rm B} = \fft12 \fft{TS}{V_{\rm th} P}\,.\label{VB}
\ee
In this formula, $(T,S)$ are the temperature and entropy of the black holes.  For Einstein's gravity, the entropy is simply one quarter of the area of the event horizon.  The quantity $P$ is the pressure associated with the cosmological constant $\Lambda$ (as in ${\cal L}=\sqrt{-g} (R - 2\Lambda)$). It is given by
\be
P=-\fft{\Lambda}{8\pi}\,.
\ee
The quantity $V_{\rm th}$ is the thermodynamical volume conjugate to $P$.  Treating the cosmological constant as a thermodynamical pressure was proposed in \cite{Kastor:2009wy,Cvetic:2010jb}.  The first law of black hole thermodynamics expands to
\be
dM=T dS + \Phi_\alpha dQ_\alpha + V_{\rm th} dP + \cdots\,,
\ee
where the dots denote contributions from further black hole hair, and the repeated $\alpha$ index implies summation.  In this picture, the mass of the black hole should be better interpreted as the enthalpy \cite{Kastor:2009wy}.  Thus we see that the butterfly velocity is related to the volume density of $TS$, which is precisely the Gibbons-Hawking surface-term \cite{Gibbons:1976ue} contribution to the action growth evaluated on the horizon \cite{Huang:2016fks}.

Assuming that the theory does not involve further dimensionful coupling constants, the first-law implies the Smarr relation
\be
M = \fft{D-2}{D-3} TS + \Phi_\alpha Q_\alpha - \fft{2}{D-3} P V_{\rm th}\,.\label{smarr}
\ee
(See \cite{Smarr:1972kt} for the original discussion of the Smarr formula in four dimensions.) For AdS planar black holes, there exists a further generalized Smarr relation, associated with the new scaling symmetry of the AdS planar black hole.  It is given by \cite{Fan:2015tua,Liu:2015tqa}
\be
M= \fft{D-2}{D-1} (TS + \Phi_\alpha Q_\alpha)\,.\label{gsmarr}
\ee
These two relations enable us to express the butterfly velocity in terms of mass, charges and their chemical potentials, namely
\bea
v_{\rm B}^2 &=& \fft{D-1}{2(D-2)} - \fft{(D-3)\Phi_\alpha Q_\alpha}{2(D-2)(2M - \Phi_\alpha Q_\alpha)}\nn\\
&=& 1 - \fft{(D-3)M}{(D-2)(2M - \Phi_\alpha Q_\alpha)}\,.\label{vb2}
\eea
(Note that when the black holes involve more hair, the Smarr relations and the above formula may involve further terms.)

In establishing (\ref{VB}), we find an identity that relates the thermodynamical volume to the Euclidean bounded volume.  This relation is purely geometrical and it depends on solely the metric functions of the near-horizon geometry.  We test this identity against a variety of static AdS black holes with both planar or spherical topologies, and find no exception.
It thus provides a simple method of calculating the thermodynamical volume even for black holes with no analytical expressions. Interestingly, if we assume that this formula is valid, the conjectured reverse isoperimetric inequality \cite{Cvetic:2010jb} follows directly from the null-energy condition, for static AdS black holes.

The paper is organized as follows:  In section 2, we first review the butterfly effects in static and isotropic AdS planar black holes.  We then obtain the universal formula (\ref{VB}) for the butterfly velocities, by making use of the geometrical identity that relates the thermodynamical and Euclidean bounded volumes.  We also address the subtleties that can arise when black holes involve non-trivial scalar hair. In section 3, we further elaborate the formula of thermodynamical volume. We also point out that it is natural to introduce black hole volume, in addition to $(T,S)$ in order to fully specify the near-horizon geometry.  In sections 4 and 5, we list a large number of AdS black holes and test the identity against their thermodynamical volumes that are derived from the first law.  In section 6, we demonstrate that the identity, together with the null-energy condition, imply the reverse isoperimetric inequality for static AdS black holes.  We then obtain a universal bound for the butterfly velocities in terms of temperature and entropy.  We conclude the paper in section 7.

\section{A universal formula for butterfly velocities}

\subsection{Shockwave and butterfly effect}

In this section, we consider a generic static AdS planar black hole in general $D$ dimensions. We focus on the isotropic configuration in the planar directions.  The metric takes the form
\be
ds_D^2 = - h(\rho)\, dt^2 + \fft{d\rho^2}{f(\rho)} + \rho^2\, d\tilde x^i d\tilde x^i\,.
\label{schwar}
\ee
In this Schwarzschild-type metric, $\rho$ and $t$ have the dimension of length, whilst $(h,f)$ and $\tilde x^i$ are dimensionless. Asymptotically at large $\rho$, the functions $(h,f)$ behave as
\be
h=g^2 \rho^2 + \cdots - \fft{\alpha}{\rho^{D-3}} + \cdots\,,\qquad
f= g^2 \rho^2 + \cdots - \fft{\tilde\alpha}{\rho^{D-3}} + \cdots\,,
\ee
where $\ell=1/g$ is the AdS radius. The mass of the black hole is proportional to the constant $\alpha$. This definition of mass is in most cases consistent with the first law.  The situation becomes more subtle where there are non-trivial scalar charges, which we shall clarify in section 2.3.

We assume that the metric describes a black hole with the event horizon located at $\rho=\rho_0>0$. We may perform Taylor expansions near the horizon
\be
h=h_1 (\rho-\rho_0) + h_2 (\rho-\rho_0)^2 + \cdots\,,\qquad
f=f_1 (\rho-\rho_0) + f_2 (\rho-\rho_0)^2 + \cdots\,.\label{hf}
\ee
In other words, $h_1=h'(\rho_0)$ and $f_1=f'(\rho_0)$.  The temperature and entropy can be calculated using the standard technique, given by
\be
T=\fft{\sqrt{h_1f_1}}{4\pi}\,,\qquad S= \ft14 \rho_0^{D-2} {\cal A}_{D-2}\,,
\ee
where ${\cal A}_{D-2}$ is the volume of the metric $d\tilde x^i d\tilde x^i$.  If this metric is compact describing a $(D-2)$-torus, we can scale the coordinate $r$ to set ${\cal A}_{D-2}$ to be any fixed value.  For example, we can choose it to be ${\cal A}_{D-2}=1$ or be the same as the volume of the unit round $S^{D-2}$, namely
\be
{\cal A}_{D-2}=\fft{2\pi^{\fft12(D-1)}}{\Gamma[\fft12(D-1)]}\,.
\ee
If $d\tilde x^i d\tilde x^i$ is noncompact like Euclidean, we can also choose ${\cal A}_{D-2}$ to be the above values, but now with the understanding that the extensive quantities such as mass, charge or entropy describe the corresponding densities.

Before proceeding, we would like to point out the fact that the near-horizon geometry is in general specified by three parameters $(h_1,f_1,\rho_0)$.  However, the well-established temperature and entropy only give two parameters, leaving the combination $h_1/f_1$ unspecified.  This demonstrates that the temperature and entropy are not enough to characterize the black hole horizon and a new quantity is called for.  We shall come back to this point in section 3.

To study the butterfly effects, it is convenient to introduce the Kruskal coordinates $(u,v)$
\be
u=e^{\kappa (\rho_* - t)}\,,\qquad v=- e^{\kappa (\rho_*+t)}\,,\qquad
\hbox{with}\qquad d\rho_* = \fft{d\rho}{\sqrt{hf}}\,.
\ee
Here $\kappa=2\pi T=\ft12\sqrt{h_1f_1}$ is the surface gravity on the horizon $\rho=\rho_0$, which corresponds to $uv=0$.  Near the horizon, we have
\bea
u v &=& (\rho-\rho_0) - \ft12 \big(\fft{f_2}{f_1} + \fft{h_2}{h_1}\big) (\rho-\rho_0)^2 + \cdots\,,\nn\\
\rho-\rho_0 &=& uv + \ft12 \big(\fft{f_2}{f_1} + \fft{h_2}{h_1}\big)\, (uv)^2 + \cdots\,.
\eea
The metric (\ref{schwar}) can now be expressed as
\be
ds_D^2 = A(u v) du dv + B(uv) d x^i d x^i\,,
\ee
where the coordinates $x^i\equiv \tilde x^i/g=\ell\, \tilde x^i$ have dimension of length, and
\be
A(u v) = \fft{1}{\kappa^2} \fft{h}{uv}\,,\qquad B(uv) = g^2 \rho^2\,.
\ee
We can expand the functions $A$ and $B$ on the horizon $uv=0$,
\be
A=A_0 + A_1\, (uv) + A_2\, (uv)^2 + \cdots\,,\qquad
B=B_0 + B_1\, (uv) + B_2\, (uv)^2 + \cdots\,.\label{AB}
\ee
It is clear that the coefficients of the Taylor expansions (\ref{hf}) and (\ref{AB}) are related. We find
\bea
A_0 &=& \fft{4}{f_1}\,,\quad A_1=\big(\fft{2f_2}{f_1} + \fft{6h_2}{h_1}\big)\fft{1}{f_1}\,,\quad
A_2=\big(\fft{3f_2^2}{4f_1^2} + \fft{f_3}{f_1} + \fft{13 f_2 h_2}{2f_1 h_1} +
\fft{19 h_2^2}{4h_1^2} + \fft{5h_3}{h_1}\big) \fft{1}{f_1}\,,\cr
B_0 &=& g^2 \rho_0^2\,,\quad B_1 = 2g^2 \rho_0\,,\quad B_2 = g^2\Big(1 + \fft{f_2}{f_1} \rho_0
+ \fft{h_2}{h_1} \rho_0\Big)\,.\label{abhf}
\eea
This allows us to translate the butterfly velocity, typically calculated using the metric of the Kruskal $(u,v)$ coordinates, in terms of variables in the more standard black hole metric of the $(t,r)$ coordinates.

The butterfly effect emerges if one releases a particle from $x=0$ on the boundary of asymptotical AdS black hole at a time $t_w$ in the past. As was described in \cite{Blake:2016wvh}, for late times (i.e. $t_w>\beta=\fft{1}{T}=2\pi/\kappa$), the energy density of this particle in Kruskal coordinates is localised on the $u=0$ horizon and it is exponentially boosted:
\be
\delta T_{uu} \sim E\,e^{\frac{2\pi}{\beta}t_w}\delta(u)\delta(\vec x)\,,
\ee
where $E$ is the initial asymptotic energy of the particle. Consequently, even the effects of an initially-small perturbation cannot be neglected and after the scrambling time $t_\ast\sim\beta\log N^2$ the back-reaction of the stress tensor on the metric becomes significant.

This results in the formation of a shockwave geometry, whose metric can then be written as \cite{Sfetsos:1994xa}
\be
ds^2 = A(uv)\,dudv + B(uv)\, dx^i dx^i - A(uv)\,\delta(u)\,h(\vec x)\,du^2.
\ee
For Einstein gravity with minimally-coupled matter, one finds that the shockwave satisfies the wave equation \cite{Sfetsos:1994xa,Blake:2016wvh,Roberts:2016wdl}
\be
(\Box-m^2)h(\vec x) \sim Ee^{\frac{2\pi}{\beta}t_w}\delta(\vec x)\label{perbein}\,.
\ee
Here $\Box$ is the Laplacian on the $(D-2)$-dimensional Euclidean metric $dx^idx^i$ and
the screening length $m$ is given by
\be
m^2 = (D-2)\frac{B_1}{A_0}\,,\label{msquare}
\ee
where $B_1$ and $A_0$ are defined by (\ref{AB}).  Thus the properties of the shockwave is encoded in the equation \eqref{perbein}. At long distances $x\gg m^{-1}$ the metric is simply given by
\be
h(\vec x) \sim \frac{Ee^{\frac{2\pi}{\beta}(t_w-t_\ast)-m|\vec x|}}{|\vec x|^{\frac{D-3}{2}}}
\ee
One can immediately read off the Lyapunov exponent $\lambda_L$ and velocity $v^{}_{\rm B}$ of these holographic theories as \cite{Roberts:2014isa}
\be
\lambda_{\rm L} = \frac{2\pi}{\beta}, \quad v^{}_{\rm B} = \frac{2\pi}{\beta m}\,.\label{lvb}
\ee

\subsection{Butterfly velocity}

As we have discussed in the introduction, we expect that the dimensionless $v_{\rm B}^{}$ may be a ratio of some thermodynamical quantities.  It follows from (\ref{lvb}) that the numerator has a factor of $T$, suggestive of a dimensionless ratio of $TS$ and the product of
another thermodynamical conjugate pair.  In fact, it follows from (\ref{lvb}), (\ref{msquare}) and (\ref{abhf}) that the butterfly velocity satisfies
\be
v_{\rm B}^2 =\sqrt{\fft{h_1}{f_1}}\, \fft{2\pi\, T}{(D-2)g^2 \rho_0} =
\fft12 \sqrt{\fft{h_1}{f_1}}\, \fft{T S}{V_{\rm E} P}\,,\label{vb3}
\ee
where
\be
P=\fft{(D-1)(D-2) g^2}{16\pi}\,,\qquad S=\ft14\rho_0^{D-2} {\cal A}_{D-2}\,,\qquad
V_{\rm E}=\fft{\rho_0^{D-1}}{D-1} {\cal A}_{D-2}\,.
\ee
Here $V_{\rm E}$ denotes the Euclidean bounded volume and its terminology will be explained in section 3.  For non-dilatonic black holes such as Schwarzschild or Ressner-Nordstr\o m (RN) metrics, one has $h=f$, hence $h_1/f_1=1$, and $V_{\rm E}$ is precisely the thermodynamical volume.  In general, we claim that the thermodynamical volume is given by
\be
V_{\rm th}=\sqrt{\fft{f_1}{h_1}}\, V_{\rm E}\,.\label{id0}
\ee
This then leads straightforwardly to the general formula (\ref{VB}) for butterfly velocities.  We shall elaborate this formula further in section 3, and then in sections 4 and 5, we shall verify this formula using a variety of AdS black holes.

In the introduction, we then made use of the the Smarr relation (\ref{smarr}) and the generalized Smarr relation (\ref{gsmarr}) to express the $v_{\rm B}^2$ as (\ref{vb2}). However, there is a subtlety when the black hole contains non-trivial scalar hair.

\subsection{Subtleties involving scalar hair}

A rather general class of AdS black holes come from Lagrangians with a scalar potential.  For simplicity, we shall consider only one scalar $\phi$, and its scalar potential is ${\cal V}(\phi)$.  The existence of the AdS vacuum of radius $\ell=1/g$ requires that ${\cal V}$ has a stationary point, say $\phi=0$, with
\be
{\cal V}'(0)=0\,,\qquad {\cal V}(0)= (D-1)(D-2) g^2\,.
\ee
Depending on the mass parameter of the scalar $\tilde m^2={\cal V}''(0)$, at large $\rho$, the scalar behaves as
\be
\phi(r)\sim \fft{\phi_1}{r^{\fft12(D-1-\sigma)}} + \fft{\phi_2}{r^{\fft12(D-1+\sigma)}} + \cdots\,,
\ee
where $\sigma$ is related to the mass parameter as $\sigma=\sqrt{4\tilde m^2g^{-2} + (D-1)^2}$. It was shown that the first law of the black hole thermodynamics takes the form \cite{Liu:2013gja,Lu:2014maa}
\be
dM=TdS + Z + \cdots\,,\label{fo}
\ee
with
\be
Z=\fft{\sigma g^2}{32\pi (D-1)} g^2
\Big((D-1+\sigma)\phi_2 d\phi_1 - (D-1-\sigma) \phi_1 d\phi_2\Big)\,.\label{z1-form}
\ee
Here the mass is given by
\be
M=\fft{(D-2) {\cal A}_{D-2}}{16\pi} \alpha\,.\label{gmass}
\ee
The above result is controversial owing to the fact that the variation of the on-shell Hamiltonian
\be
(d {\cal H})_{\rho\rightarrow\infty} = dM - Z\,,
\ee
is not integrable, and hence the energy is not well defined.  In this paper, we adopt the same strategy of \cite{Lu:2014maa}, and introduce a ``gravitational mass'' $M$ that describes the condensate of the spin-2 massless graviton.  It is then clear from the dimensional analysis that the integration constants $(\phi_1,\phi_2)$ do not involve in the Smarr relation, and hence relation (\ref{smarr}) holds even for black holes with such non-trivial scalar hair, provided that $M$ is defined as the gravitational mass, satisfying (\ref{fo}).

The first law (\ref{fo}) and the Smarr relation (\ref{smarr}) are independent of whether the AdS black hole is spherically symmetric or is of the planar type.  For the planar AdS black holes with non-trivial scalar hair, it was shown in \cite{Liu:2015tqa} that the generalized Smarr relation (\ref{gsmarr}) also works.  In other words, the scalar hair variables $(\phi_1,\phi_2)$ do not enter the Smarr nor the generalized Smarr relation.  Thus the formula (\ref{vb2}) is valid even for black holes with non-trivial scalar hair provided that $M$ is the gravitational mass defined by (\ref{gmass}).

\section{A geometrical formula for thermodynamical volumes}

In the previous section, we obtained the butterfly velocity (\ref{vb3}).  We find that it can be expressed universally as a simple ratio (\ref{VB}) of thermodynamical variables, provided that the identity (\ref{id0}) for calculating the thermodynamical volume is valid. In this section, we discuss this identity in more detail.   A static black hole metric may not always be expressed analytically in terms of the Schwarzschild-type coordinate as in (\ref{schwar}). Instead it takes a more general form
\be
ds^2 = - h(r) dt^2 + \fft{dr^2}{f(r)} + \rho(r)^2 d\Omega_{D-2,k}^2\,.
\ee
The statement of the identity (\ref{id0}) in this more general coordinate system becomes
\be
\fft{V_{\rm th}}{V_{\rm E}} = \sqrt{\fft{f}{h}}\, \fft{d\rho}{dr}\Big|_{r\rightarrow r_0}\,,
\label{id}
\ee
where $r=r_0$ is the event horizon.  It is perhaps better to express the above identity in a more abstract notation
\be
\fft{V_{\rm th}}{V_{\rm E}} = \fft{1}{\sqrt{-g_{tt}\, g_{rr}}}\, \fft{d\rho}{dr}\Big|_{r\rightarrow r_0}\,.
\label{idabs}
\ee
As explained in the introduction, the quantity $V_{\rm th}$ is the thermodynamical volume derived from the first law of black hole thermodynamics
\be
dM = T dS + \Phi_\alpha dQ_\alpha + V_{\rm th} dP + \cdots\,,\label{fl0}
\ee
where $P$ is the thermodynamical pressure
\be
P=-\fft{\Lambda}{8\pi}=\fft{(D-1)(D-2) g^2}{16\pi}\,.\label{pressure}
\ee
The quantity $V_{\rm E}$ is defined as
\be
V_{\rm E} = \fft{\rho_0^{D-1}}{D-1} {\cal A}_{D-2}\,.
\ee
where $\rho_0=\rho(r_0)$ and ${\cal A}_{D-2}$ is the volume of the metric $d\Omega_{D-2,k}^2$ which is a unit Einstein metric with
\be
\widetilde R_{ij}=(D-3) k\, \tilde g_{ij}\,,\qquad k=-1,0,1\,.
\ee
When $k=1$, and the metric is a unit round $S^{D-2}$, $V_{\rm E}$ is precisely the volume of a ball of radius $\rho_0$ in the $(D-1)$-dimensional Euclidean space.  For this reason, $V_{\rm E}$ was called Euclidean bounded volume in \cite{Cvetic:2010jb}.  Here we generalize the concept to include the $k=-1,0$ topologies as well.  Thus the identity (\ref{id}) or (\ref{idabs}) provides a purely horizon-geometric formula for calculating the thermodynamical volume.  An important advantage is that this method does not require an exact analytical expression of a black hole in order to calculate its thermodynamical volume.  By contrast, determining $V_{\rm th}$ through the first law would necessarily require that the exact solution of the black hole be known, since the first law involves both horizon quantities such as $(T,S)$ and the asymptotic quantities such as mass and charges.

    For Schwarzschild-AdS or RN-AdS black holes, one has $h=f$ in the $r=\rho$ coordinate gauge choice. The thermodynamical volume is precisely the Euclidean bounded volume.  Such black holes are typically non-dilatonic and include, for examples, \cite{DHoker:2009mmn,Bardoux:2012aw,Andrade:2013gsa,Liu:2016njg,Li:2016nll}.  For these examples, the formula (\ref{id}) is thus automatically valid.

The situations becomes more complicated when the theory involve radially-dependent scalar fields.  For example, let us consider
\be
{\cal L}=\sqrt{-g} \Big(R - \ft12 (\partial\phi)^2 - V(\phi) + \cdots\Big)\,.
\ee
The Einstein equation of motion implies that (in the $r=\rho$ gauge)
\be
\phi'^2 =- \fft{(D-2)}{\rho}\,\fft{h}{f}\,\Big(\fft{f}{h}\Big)'\,.\label{phihf}
\ee
Thus for these black holes, $V_{\rm th}\ne V_{\rm E}$, but rather their ratio satisfies the identity (\ref{id}).  In the next two sections, we shall test this identity with a variety of black holes involving scalar fields, by computing their thermodynamical volumes using the first law (\ref{fl0}).  A formula for calculating the thermodynamical volume based on the Wald formalism was also obtained in \cite{Urano:2009xn,Couch:2016exn}.  As in the case of using the first law to computing the volume, this formula of \cite{Urano:2009xn,Couch:2016exn} is useful only when one has an exact solution, since it requires an integration from the horizon to asymptotic infinity, where potential divergent terms cancel out in a intricate fashion.  This is very different from our proposed formula (\ref{id}) that depends only on the near-horizon data, and hence does not rely on the existence of an exact solution.

It is worth pointing out again that (\ref{id}) depends only on the near-horizon geometry, which does not, a priori, ``knows'' the asymptotic structure, whether it is asymptotically flat or AdS. This is suggestive that the volume $V_{\rm th}$ based on (\ref{id}) may be fundamental a quantity of black holes in general.  It may be valid even for black holes with flat (or other unusual) asymptotics, where there is no thermal pressure associated with the cosmological constant.  It happens that in asymptotic AdS black holes, where a sensible pressure can be defined, the volume $V_{\rm th}$ becomes a thermodynamical one conjugate to the pressure.  In fact, as we discussed in section 2.1, for static black holes in the $\rho=r$ gauge, the near-horizon geometry is characterized by three parameters $(h_1,f_1,r_0)$.  The entropy is a function of $r_0$, whilst $T=\sqrt{h_1 f_1}/(4\pi)$.  Thus $(T,S)$ do not specify the ratio $\sqrt{f_1/h_1}$.  Our definition of $V_{\rm th}$ precisely makes up for this ratio. It is thus natural to introduce three quantities $(T, S, V_{\rm th})$, instead of only $(T,S)$, to specify the horizon property fully.

Before finishing this section, we note that the thermalization factor which vanishes on the event horizon cancels out in $\sqrt{f/h}$.  Thus we have $\sqrt{f_1/h_1}=\sqrt{f/h}|_{r=r_0}$. It follows that the ratio of (\ref{id}) can be well defined for generic $r\ge r_0$, leading to the well-defined function
\be
\sigma(r)=\sqrt{\fft{f}{h}} \fft{d\rho}{dr}\,.
\ee
For the gauge choice $r=\rho$, it is clear that $\sigma(\infty)=1$.  In this gauge, we shall show in section 6 that $\sigma(r)$ is a monotonically-decreasing function.

\section{Testing the identity: Kaluza-Klein dyonic AdS black holes}

In this and the next section, we shall test the formula (\ref{id}) using explicit examples of AdS black holes. For non-dilatonic type of black holes, we have $h/f=1$ in the gauge choice of $r=\rho$.  In this case, it is clear that $V_{\rm th}=V_{\rm E}$ and it follows that the identity (\ref{id}) is trivially true.  Thus it is of more non-trivial to consider solutions with non-vanishing scalar fields.  Furthermore, we would also like to test the Smarr (\ref{smarr}) and generalized Smarr (\ref{gsmarr}) relations when solutions involve non-trivial scalar hair.  For almost all the exact black hole solutions  constructed in literature, the 1-form $Z$ in (\ref{z1-form}) vanishes, and the subtlety does not arise.  The only known exact solutions with non-vanishing $Z$ are the Kaluza-Klein dyonic AdS black holes \cite{Lu:2013ura} in maximal gauged supergravity and their multi-charge generalizations \cite{Chow:2013gba}.  This may be the most non-trivial example to test the identity.

\subsection{Spherically-symmetric black hole}

It was shown in \cite{Lu:2013ura} that the Lagrangian
\begin{equation}
{\cal L}=\sqrt{-g}\Big[R - \ft12 (\partial\phi)^2 - \ft14 e^{-\sqrt3\,\phi} F^2 +6g^2 \cosh\Big(\ft1{\sqrt3} \phi\Big)\Big]\,,\label{lag}
\end{equation}
admits the following static solution
\begin{eqnarray}
ds^2 &=& -(H_1 H_2)^{-\fft12} \hat f dt^2 + (H_1 H_2)^{\fft12} \Big(\fft{dr^2}{\hat f} + r^2 (d\theta^2 + \sin^2\theta\, d\varphi^2)\Big)\,,\cr
\phi&=&\fft{\sqrt{3}}{2} \log\fft{H_2}{H_1}\,,\qquad
\hat f=f_0 + g^2 r^2 H_1 H_2\,,\qquad f_0=1 - \fft{2\mu}{r}\,,\cr
A&=&\sqrt2 \Big(\fft{(1 - \beta_1 f_0)}{\sqrt{\beta_1\gamma_2}\, H_1}\, dt + 2\mu\,\gamma_2^{-1}\sqrt{\beta_2\gamma_1}\, \cos\theta\, d\varphi\Big)\,,\cr
H_1&=&\gamma_1^{-1} (1-2\beta_1 f_0 + \beta_1\beta_2 f_0^2)\,,\qquad
H_2=\gamma_2^{-1}(1 - 2\beta_2 f_0 + \beta_1\beta_2 f_0^2)\,,\cr
\gamma_1&=& 1- 2\beta_1 + \beta_1\beta_2\,,\qquad \gamma_2 = 1-2\beta_2 + \beta_1\beta_2\,.
\label{adsdyon}
\end{eqnarray}
The solution describes a dyonic AdS black hole with the horizon $r=r_0$ given by
\begin{equation}
\hat f(r_0)=1-\fft{2\mu}{r_0} + g^2 r_0^2 H_1(r_0) H_2(r_0)=0\,.
\end{equation}
The mass and electric and magnetic charges are
\bea
M= \fft{(1-\beta_1)(1-\beta_2)(1-\beta_1\beta_2) \mu}{\gamma_1 \gamma_2}\,.\qquad
Q = \fft{\mu\,\sqrt{\beta_1\,\gamma_2}}{\sqrt2\, \gamma_1}\,,\qquad
P= \fft{\mu\, \sqrt{\beta_2\, \gamma_1}}{\sqrt2\, \gamma_2}\,.
\eea
The temperature, entropy and chemical potentials are
\bea
T &=& \fft{\hat f'(r_0)}{4\pi \sqrt{H_1(r_0) H_2(r_0)}}\,,\qquad S=\pi r_0^2
\sqrt{H_1(r_0)H_2(r_0)}\,,\nn\\
\Phi_Q &=& \sqrt{\fft{2}{\beta_1\gamma_2}}
\Big(1 - \beta_1 - \fft{1-\beta_1 f_0(r_0)}{H_1(r_0)}\Big)\,,\quad
\Phi_P = \sqrt{\fft{2}{\beta_2\gamma_1}}
\Big(1 - \beta_2 - \fft{1-\beta_2 f_0(r_0)}{H_2(r_0)}\Big)\,.
\eea
The scalar hair contribution to the first law is given by
\be
Z=X dY\,,\qquad X=\fft{4g^2\mu^3(\beta_1-\beta_2)\sqrt{\beta_1\beta_2^3}}{(1-\beta_1\beta_2)
\gamma_2^2} \,,\qquad
Y=\fft{\sqrt{\beta_1}\, \gamma_2}{\sqrt{\beta_2}\, \gamma_1}\,.
\ee
The first law can then be established that \cite{Lu:2013ura}
\begin{equation}
dM=T dS + \Phi_Q\, dQ + \Phi_P\, dP + X dY +
\Upsilon d\Lambda\,,
\end{equation}
where
\begin{eqnarray}
\Upsilon&=&-\fft{r_0^3}{12\gamma_1\gamma_2} \Big(\beta_1 \beta_2 (2 \beta_1 \beta_2
-\beta_1 - \beta_2 ) \tilde f_0^3 +3 \beta_1 \beta_2 (2 - \beta_1 - \beta_2) \tilde f_0^2\cr
 &&\qquad\qquad + 3 (2 \beta_1 \beta_2-\beta_1 - \beta_2) \tilde f_0 -2 +
 \beta_1 + \beta_2\Big)\,,
\end{eqnarray}
and $\tilde f_0=1-2\mu/r_0$.  Here we also corrected some sign typos in \cite{Lu:2013ura}.  Thus for $P$ given in (\ref{pressure}), we have
\be
V_{\rm th} = -8\pi\,\Upsilon\,.
\ee
Note that the Smarr relation
\begin{equation}
M=2 T S + \Phi_Q\,  Q + \Phi_P\,  P - 2 V_{\rm th} P\,,\label{d4smarr}
\end{equation}
indeed does not involve any scalar hair.

Having obtained the thermodynamical volume, we can now test the formula (\ref{id}).  It is clear that we have
\be
\rho=r (H_1 H_2)^{\fft14}\,,\qquad V_{\rm E} = \fft{4\pi}{3} \rho(r_0)^3\,,\qquad
\fft{f}{h} = 1\,.\label{d4rhov}
\ee
It is now very simple to verify (\ref{id}).

\subsection{Toroidally-symmetric black hole}

As was emphasized before, once we have a spherically-symmetric solution, solutions with different topology of the level surfaces can be easily obtained from some simple scaling of the coordinates.  The formula (\ref{id}) is clearly invariant under such scaling and hence valid for all topologies.  However, in the toroidal limit, new scaling symmetry emerges that can lead to the generalized Smarr relation (\ref{gsmarr}).  Also in this paper, we focus on the AdS planar black holes for studying the butterfly effect. For these reasons, we discuss this example in some detail, even though the results are already implied by the previous example.  The solution is given by \cite{Lu:2013ura}
\begin{eqnarray}
ds^2 &=& -(H_1 H_2)^{-\fft12} \hat f dt^2 + (H_1 H_2)^{\fft12} \Big(\fft{dr^2}{\hat f}
+ r^2 (dx^2 + dy^2)\Big)\,,\cr
\phi&=&\ft{\sqrt{3}}{2} \log\fft{H_2}{H_1}\,,\qquad
f=- \fft{2\mu}{r} + g^2 r^2 H_1 H_2\,,\cr
A&=&\sqrt{2\mu} \Big(\fft{(r + 2\beta_1)}{\sqrt{\beta_1}\, H_1\,r}\, dt
+ 2\sqrt{\beta_2}\, xdy\Big)\,,\cr
H_1&=&1 + \fft{4\beta_1}{r} + \fft{4\beta_1\beta_2}{r^2}\,,\qquad
H_2=1 +\fft{4\beta_2}{r} + \fft{4\beta_1\beta_2}{r^2}\,.
\label{adsdyon2}
\end{eqnarray}
The horizon is located at $r=r_0$, with
\begin{equation}
\mu=\ft12 g^2 r_0^3 H_1(r_0) H_2(r_0)\,.
\end{equation}
All the thermodynamical quantities are \cite{Lu:2013ura}
\begin{eqnarray}
&&M=\mu\,,\qquad Q=\sqrt{\fft{\mu\beta_1}{2}}\,,\qquad
P=\sqrt{\fft{\mu\beta_2}{2}}\,,\cr
&&T = \fft{\hat f'(r_0)}{4\pi \sqrt{H_1(r_0) H_2(r_0)}}\,,\qquad S=\pi r_0^2
\sqrt{H_1(r_0)H_2(r_0)}\,,\nn\\
&&\Phi_Q=\fft{2\sqrt{2\mu\beta_1} (r_0 + 2\beta_2)}{r_0^2 H_1(r_0)}\,,\qquad
\Phi_P=\fft{2\sqrt{2\mu\beta_2} (r_0 + 2\beta_1)}{r_0^2 H_2(r_0)}\,,\cr
&&X = 4 g^2 (\beta_1 - \beta_2) \sqrt{\beta_1 \beta_2^3}\,,\qquad
Y=\sqrt{\fft{\beta_1}{\beta_2}}\,,\qquad P=\fft{3g^2}{8\pi}\,,\cr
&&V_{\rm th} = \ft43\pi \big(4 \beta_1 \beta_2 (\beta_1 + \beta_2) + 12 \beta_1 \beta_2 r_0
+ 3 (\beta_1 + \beta_2) r_0^2 + r_0^3\big)\,.\label{thermoquan}
\end{eqnarray}
It is now straightforward to verify that the first law is again satisfied.  The usual Smarr formula takes the same form as (\ref{d4smarr}).  In addition, there is a generalized Smarr relation
\be
M=\ft23(T S+ \Phi_Q\, Q + \Phi_P\, P)\,.
\ee
In the above discussion, we have, for convenience, assumed that the
$\R^2$ coordinates $(x,y)$ have been identified to give a 2-torus of
volume $4\pi$.  One can take any other choice for the volume, with the
understanding that the extensive quantities should be scaled by the
relative volume factor.

The quantities listed in (\ref{d4rhov}) are the same for the toroidal black hole, and then the formula (\ref{id}) can be straightforwardly shown to be true.  By utilizing both the Smarr and generalized Smarr relations, we thus have
\be
v_{\rm B}^2 = \fft34 - \fft{\Phi_Q Q + \Phi_P P}{4(2M - \Phi_Q Q - \Phi_P P)}\,.
\ee
Note that since the form of Smarr and generalized Smarr relations are identical to the dyonic RN-AdS planar black hole, it follows that the above formula applies for the dyonic RN-AdS planar black hole as well.

\section{Testing the identity: further AdS black holes}

In this section, we shall test the identity (\ref{id}) with more AdS black holes in literature that we are familiar with.  These are static black holes with spherical or more general topologies.  The list is by no means exhaustive. The presentation will also be less detailed than that in the previous section.  Instead we shall simply present the metric functions and the thermodynamical volumes in most cases.  It is worth pointing out that none of the examples in this section has non-trivial scalar hair, i.e. the 1-form $Z$ in (\ref{z1-form}) vanishes for all.

\subsection{R-charged black holes in gauged supergravities}

The thermodynamical quantities including the volumes for R-charged black holes in gauged supergravities in $D=4,5$ and 7 dimensions were obtained in \cite{Cvetic:2010jb}.  We find that these formulae are particularly useful for our task of verifying the identity (\ref{id}).

\subsubsection{$D=4$ four-charge black hole}

This solution was constructed in \cite{Duff:1999gh} and the metric functions are
\bea
h&=&f=\prod_{i=1}^n H_i^{-\fft12} \hat f\,,\qquad \rho=r \prod_{i=1}^4 H_i^{\fft14}\,,\nn\\
\hat f&=&1 - \fft{2m}{r} + g^2 r^2 \prod_{i=1}^4 H_i\,,\qquad H_i =1 + \fft{q_i}{r}\,.
\eea
The thermodynamical and Euclidean volumes are \cite{Cvetic:2010jb}
\be
V_{\rm th} = \ft13\pi r_0^3 \prod_{i=1}^4 H_i(r_0) \sum_{j=1}^4 \fft{1}{H_j(r_0)}\,,\qquad
V_{\rm E} = \ft43\pi \rho(r_0)^3\,.
\ee
Since the thermalization factor $\hat f$ cancels in $f/h$, it follows that in the actual verification of (\ref{id}), $r_0$ can be treated as if it is a generic $r$.  This avoids the precise determination of $r_0$ which at times can be tedious. This is the case for all examples.

\subsubsection{$D=5$ three-charge black hole}

The solution was constructed in \cite{Behrndt:1998jd}.  The metric functions
are
\bea
h&=&\prod_{i=1}^3 H_i^{-\fft23}\, \hat f\,,\qquad f=\prod_{i=1}^3 H_i^{-\fft13}\, \hat f\,,\qquad \rho=r \prod_{i=1}^3 H_i^{\fft16}\,,\nn\\
\hat f&=&1 - \fft{2m}{r^2} + g^2 r^2 \prod_{i=1}^3 H_i\,,\qquad H_i =1 + \fft{q_i}{r^2}\,.
\eea
The thermodynamical and Euclidean volumes are \cite{Cvetic:2010jb}
\be
V_{\rm th}= \ft16 \pi^2 r_0^4\,\prod_{i=1}^3 H_i(r_0)\, \sum_{j=1}^3 \fft1{H_j(r_0)}\,,\qquad
V_{\rm E}=\ft12 \pi^2 \rho(r_0)^4\,.
\ee

\subsubsection{$D=6$ two-charge black hole}

This solution can be extracted from a general class of solutions obtained in \cite{Liu:2012ed}. The metric functions are
\bea
h &=& (H_1 H_2)^{-\fft34}\hat f\,,\qquad f= (H_1 H_2)^{-\fft14} \hat f\,,\qquad
\rho=r (H_1 H_2)^{\fft18}\,,\nn\\
\hat f &=& 1 - \fft{2m}{r^3} + g^2 r^2 H_1 H_2\,,\qquad H_i = 1 + \fft{q_i}{r^3}\,.
\eea
We find that the thermodynamical and Euclidean volumes are
\be
V_{\rm th} = \ft1{15}\pi^2 r_0^5 \Big(2 H_1(r_0) H_2(r_0) + 3 H_1(r_0) + 3 H_2(r_0)\Big)\,,
\qquad V_{\rm E}=\ft8{15}\pi^2 \rho(r_0)^5\,.
\ee

\subsubsection{$D=7$ two-charge black hole}

The solution was obtained in \cite{Cvetic:1999xp}.  The metric functions are
\bea
h &=& (H_1 H_2) ^{-\fft45}\,\hat f\,,\qquad f=(H_1 H_2)^{-\fft15}\, \hat f\,,\qquad
\rho = r (H_1 H_1)^{\fft1{10}}\,,\nn\\
\hat f &=& 1 - \fft{2m}{r^4} + g^2 r^2 H_1 H_2\,,\qquad H_i = 1 + \fft{q_i}{r^4}\,,
\eea
The thermodynamical and Euclidean volumes are \cite{Cvetic:2010jb}
\be
V_{\rm th} = \fft{\pi^3}{30} r_0^6  \Big(H_1(r_0) H_2(r_0) + 2H_1 (r_0) + 2 H_2 (r_0)\Big)\,,\qquad V_{\rm E}= \fft{\pi^3}{6} \rho(r_0)^6\,.
\ee
For all above $D=4,5,6,7$ examples, it is simple algebra to verify (\ref{id}).

\subsection{Charged dilatonic AdS black holes}

A general class of Lagrangians involving a minimally-coupled dilaton and two Maxwell fields with some appropriate scalar potential were proposed \cite{Lu:2013eoa}:
\bea
e^{-1} {\cal L}_D &=& R - \ft12 (\partial\phi)^2 - \ft14 e^{a_1\phi} F_1^2 - \ft14 e^{a_2\phi} F_2^2 - V(\phi)\,,\nn\\
V &=& \big(\fft{dW}{d\phi}\big)^2 - \fft{D-1}{2(D-2)} W^2\,,\nn\\
W&=&\fft1{\sqrt2} N_1 (D-3) g \big(e^{-\fft12 a_1 \phi} - \fft{a_1}{a_2} e^{-\ft12 a_2\phi}\big)\,,
\eea
where dilatonic coupling constants $(a_1,a_2)$ satisfy
\be
a_1 a_2 = - \fft{2(D-3)}{D-2}\,.\label{a1a2}
\ee
One may introduce a pair of constants $(N_1,N_2)$, defined by
\be
a_1 = \fft{4}{N_1} - \fft{2(D-3)}{D-2}\,,\qquad a_2^2 = \fft{4}{N_2} - \fft{2 (D-3)}{D-2}\,.
\ee
The constraint (\ref{a1a2}) implies that
\be
N_1 a_1 + N_2 a_2 = 0\,,\qquad N_1 + N_2 = \fft{2(D-2)}{D-3}\,.\label{n1n2}
\ee
The charged AdS black hole is given by \cite{Lu:2013eoa}
\bea
ds^2 &=& - (H_1^{N_1} H_2^{N_2})^{-\fft{D-3}{D-2}} \, \hat f\, dt^2 +
 (H_1^{N_1} H_2^{N_2})^{\fft{1}{D-2}}\big(\hat f^{-1} dr^2 + r^2 d\Omega_{D-2}^2\big)\,,\nn\\
A_i &=& \sqrt{\ft{N_1(\mu+q_i)}{q_i}} H_i^{-1} dt\,,\qquad
\phi= \ft12 N_1 a_1 \log H_1 + \ft12 N_2 a_2 \log H_2\,,\nn\\
f &=& 1 - \fft{\mu}{r^{D-3}} + g^2 r^2 H_1^{N_1} H_2^{N_2}\,,\qquad H_i = 1 + \fft{q_i}{r^{D-3}}\,.
\eea
The solution has three integration constants $(\mu, q_1, q_2)$, parameterizing the mass and two electric charges.  The thermodynamical quantities are
\bea
M &=& \fft{{\cal A}_{D-2}}{16\pi}\big( (D-2)\mu + (D-3) (N_1 q_1 + N_2 q_2)\big)\,,\nn\\
T &=& \fft{\hat f'}{4\pi \sqrt{H_1^{N_1} H_2^{N_2}}}\,,\qquad S=\ft14 {\cal A}_{D-2} \, \rho^{D-2}\,,\nn\\
\Phi_i &=& \sqrt{\ft{N_i(\mu+q_i)}{q_i}} \big(1 - \fft{1}{H_i}\big)\,,\qquad
Q_i= \fft{(D-3) {\cal A}_{D-2}}{16\pi} \sqrt{N_i q_i (\mu+ q_i)}\,,\nn\\
V_{\rm th} &=& \fft{{\cal A}_{D-2} r_0^{D-1} H_1^{N_1-1} H_2^{N_2-1}}{2(D-1)(D-2)}
\Big( (D-3) (N_2 H_1 + N_1 H_2)\nn\\
&& +\big[2(D-2) - (D-3) (N_1 + N_2)\big] H_1 H_2\Big)\,,\qquad P= \fft{(D-1)(D-2) g^2}{16\pi}\,.\label{dilvth}
\eea
These quantities satisfy the first law
\be
dM= T dS + \Phi_1 dQ_1 + \Phi_2 dQ_2 + V_{\rm th} dP\,.
\ee
The Euclidean volume on the other hand is given by
\be
V_E = \fft{{\cal A}_{D-2}}{D-1} \rho^{D-1}\Big|_{r=r_0}\,,\qquad \rho=r\,
(H_1^{N_1} H_2^{N_2})^{\fft{1}{2(D-2)}}\,.
\ee
The identity (\ref{id}) is then easy to verify since
\be
\fft{f}{h}=\big(H_1^{N_1} H_2^{N_2}\big)^{-\fft{D-4}{D-2}}\,.
\ee
It is of interest to note that the second condition in (\ref{n1n2}) implies that the square-bracket term of $V_{\rm th}$ in (\ref{dilvth}) actually vanishes.  However, if we include this, then both the first law and the identity (\ref{id}) are valid for generic $(N_1,N_2)$ without the constraints (\ref{dilvth}).

In \cite{Chow:2011fh}, analogous Lagrangian inspired by bosonic string theory were proposed. Both static and rotating (with single rotation) AdS black holes were constructed. The metric functions for the static black holes are \cite{Chow:2011fh}
\bea
h &=& (H_1 H_2)^{-\fft{D-3}{D-2}}\,\hat f\,,\qquad f=(H_1 H_2)^{-\fft{1}{D-2}}\, \hat f\,,\qquad \rho=r (H_1 H_2)^{\fft{1}{2(D-2)}}\,,\nn\\
\hat f &=& 1  - \fft{2m}{r^{D-3}} + g^2 r^2 H_1 H_2\,,\qquad H_i = 1 + \fft{q_i}{r^{D-3}}\,.
\eea
We find that thermodynamical and Euclidean volumes are
\bea
V_{\rm th} &=& \fft{{\cal A}_{D-2}}{2(D-1)(D-2)} r_0^{D-1} (H_1 H_2)^{-\fft{D-4}{D-2}}
\Big(2 H_1 H_2 + (D-3) (H_1 + H_2)\Big)\,,\nn\\
V_{\rm E} &=& \fft{{\cal A}_{D-2}}{D-1}\,\rho(r_0)^{D-1}\,.
\eea
Note that $V_{\rm th}$ is the same as (\ref{dilvth}) with $N_1=N_2=1$.

\subsection{Scalar hairy AdS black holes}

Large classes of scalar hairy AdS black holes were constructed in \cite{Anabalon:2012ta,
Anabalon:2013qua, Anabalon:2013sra,Gonzalez:2013aca,Acena:2013jya,Feng:2013tza,Fan:2015tua}.  In all these examples, there is no non-trivial scalar hair and hence the thermodynamical quantities and the first law can be completely determined by the metric functions alone, without needing to know the solution of the scalar field.
Here we consider the examples of \cite{Feng:2013tza}.  The metric functions are
\bea
h \!\!&=&\!\! - H^{-1-\mu} \hat f\,,\qquad f=H^{-\fft{1+\mu}{D-3}} \hat f\,,\qquad
\rho=r H^{\fft{1+\mu}{2(D-3)}}\,,\qquad H=1 + \fft{q}{r^{D-3}}\nn\\
\hat f\!\! &=&\!\! H + g^2 r^2 H^{\fft{D-2}{D-3} (1+\mu)} -
\alpha^2 r^2 (H-1)^{\fft{D-1}{D-2}}\, {}_2F_1[1, \ft{D-2}{D-3} (1+\mu);
\ft{2(D-2)}{D-3}; 1 - H^{-1}]\,,
\eea
We find that the thermodynamical quantities are
\bea
M &=& \fft{(D-2){\cal A}_{D-2}}{16\pi} q \Big(\mu + \alpha q^{\fft{2}{D-3}}\Big)\,,\qquad
P= \fft{(D-1)(D-2)g^2}{16\pi}\,,\nn\\
T &=& \fft{f'}{4\pi} H^{-\fft{(D-2)(1+\mu)}{2(D-3)}}\,,\qquad
S=\fft14 r_0^{D-2} {\cal A}_{D-2} H^{\fft{(D-2)(1+\mu)}{2(D-3)}}\,,\nn\\
V_{\rm th} &=& \fft{{\cal A}_{D-2}}{2(D-1)} r_0^{D-1} H^{\fft{(D-2)(1+\mu)}{D-3}}\Big(
1-\mu + (1+\mu) H^{-1}\Big)\,.
\eea

In \cite{Fan:2015tua} two more classes of AdS planar black holes were obtained.  In the first class, the metric functions are
\be
h=f=r(r+q) \Big(g^2 -\alpha \big(\fft{q}{r}\big)^{D-1}\, {}_2 F_1[D-1, \ft12 D; D; -\ft{q}{r}]\Big)\,,\qquad \rho=\sqrt{r(r+q)}\,.
\ee
The thermodynamical quantities including the volume are
\bea
M &=& \fft{(D-2){\cal A}_{D-2}}{16\pi} \alpha q^{D-1}\,,\qquad
V_{\rm th} = \fft{{\cal A}_{D-2}(2r_0+q)}{2(D-1)} \big(r_0(r_0+q)\big)^{\fft12 D-1}\,,\nn\\
T &=& \ft1{4\pi} (D-1)\alpha q^{D-1} \big(r_0(r_0+q)\big)^{1-\fft12 D}\,,\qquad
S=\ft14 \big(r_0(r_0+q)\big)^{\fft12 D-1}{\cal A}_{D-2}\,,
\eea
In the second class, the metric functions are
\bea
h&=& r^2 \Big(g^2 + \alpha \nu \lambda^{-2\nu} \big(\Gamma(\nu, \lambda^2 \phi^2) - \Gamma(\nu)\big)\Big)\,,\qquad f=e^{2\lambda^2 \phi^2}\,h\,,\qquad \rho=r\,,\nn\\
\nu &=& \fft{D-1}{2\mu}\,,\qquad
\lambda^2 = \fft{\mu}{4(D-2)}\,,\qquad \phi=\big(\fft{q}{r}\big)^\mu\,.
\eea
The mass, temperature, entropy and the volume are
\bea
M &=& \fft{(D-2){\cal A}_{D-2}}{16\pi} \alpha q^{D-1}\,,\qquad T= \fft{\alpha (D-1) q^{D-1}}{4\pi r_0^{D-2}}\,,\qquad S=\ft14 {\cal A}_{D-2} r_0^{D-2} \,,\nn\\
V_{\rm th} &=& \fft{{\cal A}_{D-2} r_0^{D-1}}{D-1} \exp\big(\fft{\mu q^{2\mu}}{4(D-2) r_0^{2\mu}}\big)=V_{\rm E}\,  \exp\big(\fft{\mu q^{2\mu}}{4(D-2) r_0^{2\mu}}\big) \,.
\eea

To summarize, with the data provided, the formula (\ref{id}) can be successfully verified against all of the examples studied in this section.

\section{Reverse isoperimetric inequality and the $v^{}_{\rm B}$ bound}

In the previous two sections, we have tested a large number of AdS black holes and verified the geometric formula (\ref{id}) for the black hole thermodynamical volume.  In \cite{Cvetic:2010jb}, a conjecture of reverse isoperimetric inequality was proposed after having examined the relation of the thermodynamical volume and the entropy for a large number of AdS static and rotating black holes.  For static black holes, the inequality can be equally stated as
\be
V_{\rm th}\ge V_{\rm E}\,.\label{ineq}
\ee
Physically, this inequality implies that for a given amount of entropy, Schwarzschild black hole requires least space volume to store them.  We now show that if we assume the identity (\ref{id}), the above inequality is a natural consequence of the null-energy condition in Einstein's theory.

For convenience, we would like to choose a coordinate gauge $\rho=r$, and hence the metric is
\be
ds^2_D= - h(r) dt^2 + \fft{dr^2}{f(r)} + r^2 d\Omega_{D-2,k}^2\,,\qquad k=-1,0,1\,.
\ee
It follows from (\ref{id}) that the inequality is a consequence of
\be
\sqrt{\fft{f}{h}}\,\Big|_{r=r_0}\ge 1\,.\label{h/f}
\ee

To proceed, it is convenient to define a vielbein base
\be
e^{\bar 0} = \sqrt{h} dt\,,\qquad e^{\bar 1} = \fft{dr}{\sqrt{f}}\,,\qquad
e^{\bar i} = r \tilde e^{\bar i}\,,\qquad i=2,3,\ldots,D-1\,,
\ee
where the barred indices denote those in the tangent spacetime. following from the Einstein equation $T^{ab}=G^{ab}$, we find the non-vanishing components of the matter energy-momentum tensor
\bea
T^{\bar 0\bar 0} &=& - \fft{(D-2)(D-3)(f - k)}{2r^2} - \fft{(D-2) f'}{2r}\,,\nn\\
T^{\bar 1\bar 1} &=& \fft{(D-2)(D-3)(f - k)}{2r^2} + \fft{(D-2) f h'}{2rh}\,,\\
T^{\bar i\bar j} &=& \Big(\fft{(D-3)\big((D-4)(f-k)+rf'\big)}{2r^2} -
\fft{fh'^2}{4h^2} + \fft{\big(2(D-3)f + rf'\big)h'}{4rh} +
\fft{f h''}{2h}\Big)\delta^{ij}\,.\nn
\eea
The null-energy condition implies that
\be
T^{\bar 0\bar0} + T^{\bar 1\bar 1}=- \fft{(D-2)h}{2r} \Big(\fft{f}{h}\Big)'\ge 0\,.\label{nulle}
\ee
(Compare this with (\ref{phihf}).) Since the function $h$ and coordinate $r$ are non-negative from the horizon to asymptotic infinity, it follows that $(f/h)'\le 0$.  In other words, $f/h$ is monotonically decreasing.  It is conventional to scale the time coordinate such that $f/h=1$ at the asymptotic infinity.  It follows that the statement (\ref{h/f}) must be true and thus we have proven the inequality (\ref{ineq}) for static AdS black holes.

Applying the inequality (\ref{ineq}) to the butterfly velocity (\ref{VB}), we have
\be
v_{\rm B}^2 \le \fft{TS}{2V_{\rm E}\, P}= \fft{2\pi T}{(D-2) \rho_0 g^2}\,,\label{bound1}
\ee
where we have used
\be
S=\ft14 {\cal A}_{D-2}\, \rho_0^{D-2}\,,\qquad V_{\rm E}=\fft{{\cal A}_{D-2}}{D-1}\, \rho_0^{D-1}\,.
\ee
The bound (\ref{bound1}) can be better expressed as
\be
v_{\rm B}^2\, \fft{(4s)^{\fft{1}{D-2}}}{T\ell} \le \fft{2\pi}{D-2}\,,\label{bound2}
\ee
where $s=S/(\ell^{D-2} {\cal A}_{D-2})$ is the (dimensionless) entropy density and $\ell=1/g$ is the AdS radius. Thus we see that for fixed entropy, the upper bound for the velocity is proportional to $\sqrt{T}$. It is worth pointing out that the essence of the bound (\ref{bound2}) is (\ref{h/f}) and hence the bound can be proven without making use the reverse isoperimetric inequality.  Nevertheless, our approach shows that the bound and the inequality are closely related.

\section{Conclusions}

In this paper, we studied the butterfly effect of static and isotropic AdS planar black holes in the framework of Einstein's theory of general relativity, with only minimally-coupled matter.  We found a purely horizon-geometric formula (\ref{id}) to calculate the thermodynamic volumes of the general static AdS black holes.  This allows us to calculate the volume even for the cases with no exact analytical solutions. We verified this identity for a large number of black holes, with no exception.  Since the near-horizon geometry of a black hole does not, a priori, ``knows'' the asymptotic structure, it follows that the volume based on (\ref{id}) may be fundamental and it may be valid even for black holes with flat (or other unusual) asymptotics, where there is no thermal pressure associated with the cosmological constant.  Indeed, the near-horizon geometry of a static black hole involves three parameters, and we showed that they could be fully specified by $(T,S, V_{\rm th})$. That black hole volume might be as fundamental as the horizon area is both intriguing and perplexing since it contradicts the well-established holographic nature of black holes.

We found two applications of (\ref{id}) in this paper.  One is that the butterfly velocity can now be expressed universally as (\ref{VB}).  In other words, it is related to the volume density of the quantity $TS$, which is precisely the contribution to the action growth from the Gibbons-Hawking surface term evaluated on the horizon \cite{Huang:2016fks}. By utilizing the Smarr and generalized Smarr relations, the butterfly velocities can also be expressed in terms of conserved quantities including mass, charges and their chemical potentials, etc.  We derived our results for isotropic solutions. For anisotropic ones, it is reasonable to believe that (\ref{VB}) may be still valid, whilst the generalized Smarr relation will alter if it still exists.

Another application is that if we assume this identity, the conjectured reverse isoperimetric inequality proposed in \cite{Cvetic:2010jb} then follows straightforwardly from the null-energy condition for the static AdS black holes.  This may point a possible direction of proving the conjecture in general. (It should be mentioned that violations of the inequality were found for some somewhat exotic black holes, see e.g.~\cite{Hennigar:2014cfa,Hennigar:2015cja}.)

We also found that the reverse isoperimetric inequality was in one-to-one correspondence to an upper bound for the butterfly velocities in AdS planar black holes.  It is fascinating to see that the purely geometric inequality is related to the bound of a physical process.

\section*{Acknowledgement}

We are grateful to Chris Pope and Hong-Bao Zhang for discussions. The work is supported in part by NSFC grants NO. 11475024, NO. 11175269 and NO. 11235003.

\end{document}